\documentclass[%
 reprint,superscriptaddress,
 amsmath,amssymb,
 aps,prb,
]{revtex4-2}
\usepackage{graphicx}
\usepackage{dcolumn}
\usepackage{bm}
\usepackage{float}

\usepackage{siunitx}
\newcommand{\RNum}[1]{\uppercase\expandafter{\romannumeral #1\relax}}

\begin{document}
\preprint{APS/123-QED}
 
\makeatletter
\renewcommand*{\@fnsymbol}[1]{\ensuremath{\ifcase#1\or \mathsection\or *\or \dagger\or
    \ddagger\or \mathparagraph\or \|\or **\or \dagger\dagger
    \or \ddagger\ddagger \else\@ctrerr\fi}}
\makeatother

\title{Strong coupling and interfering resonances in isolated van der Waals nanoresonators }

\author{Qi Ding}
\affiliation{College of Physics, Sichuan University, Chengdu 610064, China}
\author{Swain Ashutosh}
    \affiliation{Nonlinear Physics Center, Research School of Physics, Australian National University, Canberra ACT 2601, Australia}
    \affiliation{School of Physics and Astronomy, University of Glasgow, Glasgow, G12 8QQ, UK}
\author{Luca Sortino}
    \affiliation{Chair in Hybrid Nanosystems, Nano-Institute Munich, Faculty of Physics, Ludwig-Maximilians-Universit\"{a}t M\"{u}nchen, 80539 Munich, Germany}
\author{Thomas Weber}
    \affiliation{Chair in Hybrid Nanosystems, Nano-Institute Munich, Faculty of Physics, Ludwig-Maximilians-Universit\"{a}t M\"{u}nchen, 80539 Munich, Germany}
\author{Lucca K\"{u}hner}
    \affiliation{Chair in Hybrid Nanosystems, Nano-Institute Munich, Faculty of Physics, Ludwig-Maximilians-Universit\"{a}t M\"{u}nchen, 80539 Munich, Germany}
\author{Stefan A Maier}
    \affiliation{School of Physics and Astronomy, Monash University Clayton Campus, Melbourne, Victoria 3800, Australia}
    \affiliation{Department of Physics, Imperial College London, London SW7 2AZ,UK}
\author{Sergey Kruk}
    \affiliation{Nonlinear Physics Center, Research School of Physics, Australian National University, Canberra ACT 2601, Australia}
\author{Yuri Kivshar}
    \affiliation{Nonlinear Physics Center, Research School of Physics, Australian National University, Canberra ACT 2601, Australia}
    \email{yuri.kivshar@anu.edu.au}
\author{Andreas Tittl}
    \affiliation{Chair in Hybrid Nanosystems, Nano-Institute Munich, Faculty of Physics, Ludwig-Maximilians-Universit\"{a}t M\"{u}nchen, 80539 Munich, Germany}
    \email{andreas.tittl@physik.uni-muenchen.de}
   \author{Wei Wang}
\affiliation{College of Physics, Sichuan University, Chengdu 610064, China} 
\email{w.wang@scu.edu.au}

\date{\today}

\begin{abstract}
The study of strong light-matter interaction in van der Waals materials is at the forefront of current research in physics and chemistry, and it can be enhanced dramatically by employing resonances. Here we present the first observation of quasi-bound states in the continuum (qBICs) realized via polaritonic interfering resonances in isolated WS$_2$ nanodisks. We experimentally validate the existence of polaritonic qBICs driven by intrinsic coupling of Mie resonances and excitons. The system exhibits exceptionally strong light-matter interaction with a measured Rabi splitting exceeding 310 meV - the largest reported value among all transition metal dichalcogenide (TMDC) self-hybridized systems to date. The giant coupling strength stems from qBIC-induced in-plane field enhancement, which strongly interacts with in-plane excitonic dipoles while suppressing radiative losses. Polarization-controlled measurements further demonstrate selective excitation of qBIC through switching incident polarization to specific orthogonal configurations. The observed polarization-dependent coupling provides an additional degree of freedom to control over the hybrid states' spectral characteristics and spatial field distributions. Our demonstrations provide a pathway for engineering high-quality light-matter hybrid states in compact nanostructures, with potential applications in on-chip photonics, polaritonics, and quantum optics.

\end{abstract}

\pacs{Valid PACS appear here}
\maketitle

\section{Introduction}

The pursuit of strong light-matter interaction in nanophotonic architectures lies at the heart of polariton physics, where hybridized photonic-electronic states unlock exciting phenomena ranging from Bose-Einstein condensation to low-threshold polariton lasers.\cite{Hertzog2019CSR} Central to this endeavor are transition metal dichalcogenides (TMDCs), which are van der Waals materials hosting tightly bound excitons with exceptional room-temperature stability (binding energies $>$ 200 meV). Their unique valleytronic properties, combined with tunability via strain, dielectric environment, and electrostatic gating, position TMDCs as versatile platforms for the study of strong light-matter interactions at the nanoscale.\cite{Andergachew24OEA} Critically, TMDCs maintain large excitonic oscillator strengths even in ultrathin configurations (10-100 nm), enabling direct integration with planar photonic systems while preserving quantum confinement effects.\cite{Xie2022JOP}

Realizing strong coupling necessitates concurrent optimization of three parameters: (1) deep subwavelength light confinement to enhance local field intensities, (2) spectral overlap between optical resonances and excitonic transitions, and (3) spatial mode matching to maximize coupling strengths. While these conditions have been met in TMDC monolayers coupled to plasmonic nanostructures, photonic crystals, and microcavities, such hybrid systems face intrinsic limitations. Plasmonic platforms suffer from ohmic losses restricting Q-factors, whereas multilayer microcavities demand precise alignment and lack post-fabrication tunability. \cite{Shasha24OES}

Recent advances leveraging photonic bound states in the continuum (BICs) address these challenges by offering lossless dielectric confinement with ultrahigh Q-factors and geometrically tunable resonances. \cite{Hsu16NRM,Wang25NL,Junxing23OES,Lin2025NP,Luca2025NP}
Current implementations of TMDC-BIC coupling, however, rely on heterogeneous integration-mechanically transferring exfoliated TMDC layers onto prefabricated all-dielectric metasurfaces (e.g., Si, TiO$_2$). This approach introduces interfacial defects, strain-induced exciton shifts, and reduced coupling efficiency due to a limited overlap between evanescent metasurface fields and TMDC excitons. The insertion of protective buffer layers like h-BN, while mitigating environmental degradation, further diminishes field-exciton interaction strength. These constraints highlight the need for monolithic systems that intrinsically unite exciton materials with photonic resonators.

Self-hybridized architectures, where the active material itself forms the photonic cavity, present an elegant solution by eliminating interfacial losses and enabling direct modal overlap.\cite{Munkhbat18ACSP,Verre19NN,Shen22NC,Zotev23LPR,Weber23NM,You23OL,Ding24OL,Shen24PRB1,Shen24PRB2,Zhou24PRB,Yue25PRB} Pioneering work demonstrated this concept in tungsten disulfide (WS$_2$) metasurfaces supporting symmetry-protected BICs, achieving Rabi splittings of over 110 meV.\cite{Weber23NM} Nevertheless, extending this paradigm to standalone subwavelength resonators – a critical step toward scalable quantum emitters and on-chip polaritonic circuits – remains unexplored. Recent theoretical insights suggest that individual dielectric nanoparticles can support quasi-BIC (qBIC) modes through interference between optical modes, combining subwavelength confinement with ultra-high Q factors in lossless systems.\cite{Rybin17PRL,Bogdanov19AP,Mercade20OL,Odit20AM,Huang21AP,Nurrahman23OL} While such isolated quasi-BIC resonators have shown enhanced light-matter interactions\cite{Ding2025APL,Mylnikov20ACSN,Jia22CPB,Liu22PR} and nonlinear processes,\cite{Koshelev2020Science,Anastasiia23SA,Carletti19PRR,Volkovskaya20Nanophotonics,Panmai22NC} their potential for strong light-matter interactions in self-hybridized configurations has not yet been realized.

In this paper, we demonstrate polaritonic qBICs in isolated WS$_2$ nanodisks via intrinsic coupling of Mie resonances and excitons. Theoretical modeling reveals that interference between two vectorial Mie modes generates photonic qBICs, which hybridize with in-plane excitons to produce anti-crossing features with 310 meV Rabi splitting — the largest reported in self-hybridized TMDC systems. Experimentally, azimuthally and TE-polarized excitations confirm the existence of polaritonic qBICs. We further reveal that the observed giant energy splitting originates from qBIC-induced in-plane field enhancement, which strongly interacts with excitonic dipoles while suppressing radiative losses. Polarization control enables selective activation of qBIC, allowing the modulation of hybrid states’ spectral properties. We believe that our findings provide a general strategy for engineering high-quality light-matter hybrid states in compact nanostructures, paving the way for advances in nanoscale light manipulation, energy transfer, and sensing technologies.

\section{Results and Discussion}

\begin{figure*}[t]
  \includegraphics[width=0.98\linewidth]{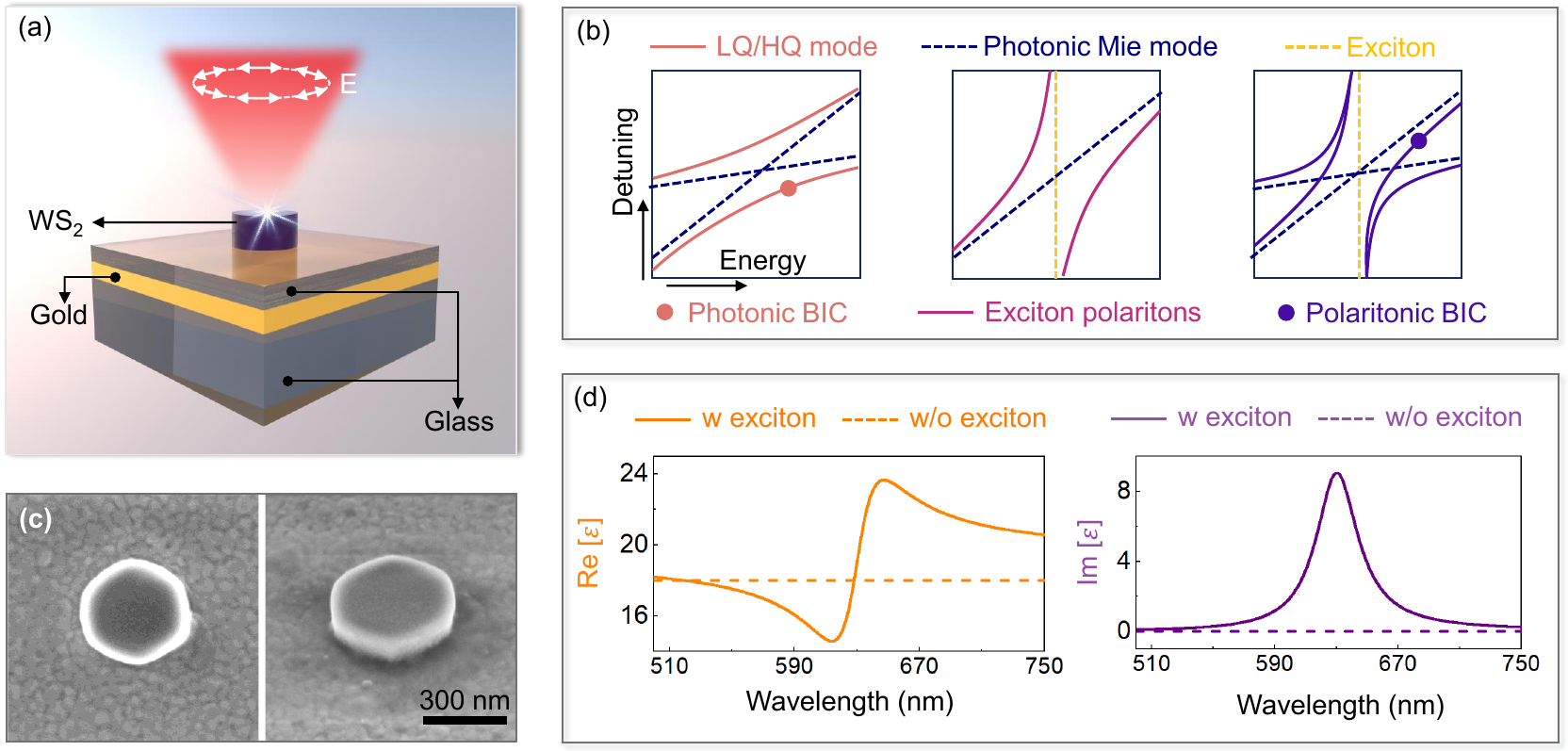}
  \caption{ WS$_2$ nanoresonators. (a) Sketch of a bulk WS$_2$ nanodisk on a substrate incident by azimuthal wave. (b) Concept of photonic BIC (left penel), exciton polaritons (middle panel) and polaritonic BIC (right panel). (c) False-colour top-view and tilt-view scanning electron microscope (SEM) micrographs of fabricated WS$_2$ nanodisks. (d) The artificial real part (left panel) and the imaginary part (right panel) of the in-plane complex permittivity with and without the exciton for WS$_2$.}
  \label{fig1}
\end{figure*}

The designed system, as shown in Fig.~\ref{fig1}(a), consists of a WS$_2$ nanodisk placed on a multilayered substrate. The substrate is composed of a 55 nm SiO$_2$ spacer, a 100 nm-thick gold reflector, and a SiO$_2$ base layer. As a high-refractive-index nanoresonator, the WS$_2$ nanodisk can support two vectorial Mie modes when illuminated with an azimuthally polarized beam at normal incidence. Here, the gold film plays a crucial role in tuning the phase difference between the vectorial Mie modes supported by the nanodisk~\cite{Koshelev2020Science}. Interference between these Mie modes leads to the formation of a Friedrich–Wintgen BIC (FWBIC)\cite{Odit20AM}, as illustrated in Fig.~\ref{fig1}(b) (left panel).  The excited FWBIC mode further couples with the intrinsic A exciton of the bulk WS$_2$ nanodisk, leading to the formation of an excitonic BIC polariton, as depicted in Fig.~\ref{fig1}(b) (right panel). Compared with traditional exciton polaritons formed by the coupling of optical modes in nanostructures with excitons in semiconductor materials (middle panel), the polaritonic BIC in this system emerges through the interference of two vectorial Mie modes, forming a qBIC, which subsequently couples with the intrinsic excitons. One key advantage of this self-hybridized approach is that the qBIC mode is confined within a highly compact isolated nanoresonator, exhibiting extremely strong subwavelength field confinement. This enables highly efficient coupling with the intrinsic exciton, significantly enhancing the strength of light-matter interactions. 

In our experiments, we fabricated a series of WS$_2$ nanodisks with different radii using a combination of electron beam lithography (EBL), chemical vapor deposition (CVD) and dry etching techniques (see Methods for details). Scanning electron microscopy (SEM) images of the nanodisk are shown in Fig.~\ref{fig1}(c) (see also Section 1, Supporting Information).

To investigate the optical properties of the WS$_2$ nanodisk system, we performed full-wave simulations using commercial simulation software (COMSOL Multiphysics). In our simulation, we used the dielectric function of WS$_2$ obtained from experimental ellipsometry measurements. The real (solid line, left panel) and imaginary (solid line, right panel) parts of the dielectric function are shown in Fig.~\ref{fig1}(d). 

We first consider the nanodisk as a pure high-refractive-index nanoresonator by disabling the A-exciton contribution. For this purpose, we fitted the dielectric function by a sum of Lorentzian oscillators and excluded the contribution of the A exciton (see Supporting Information for details).  The calculated background dielectric function is shown in Fig.~\ref{fig1}(d) (orange dashed line).  

\begin{figure*}[t]
  \includegraphics[width=0.95\linewidth]{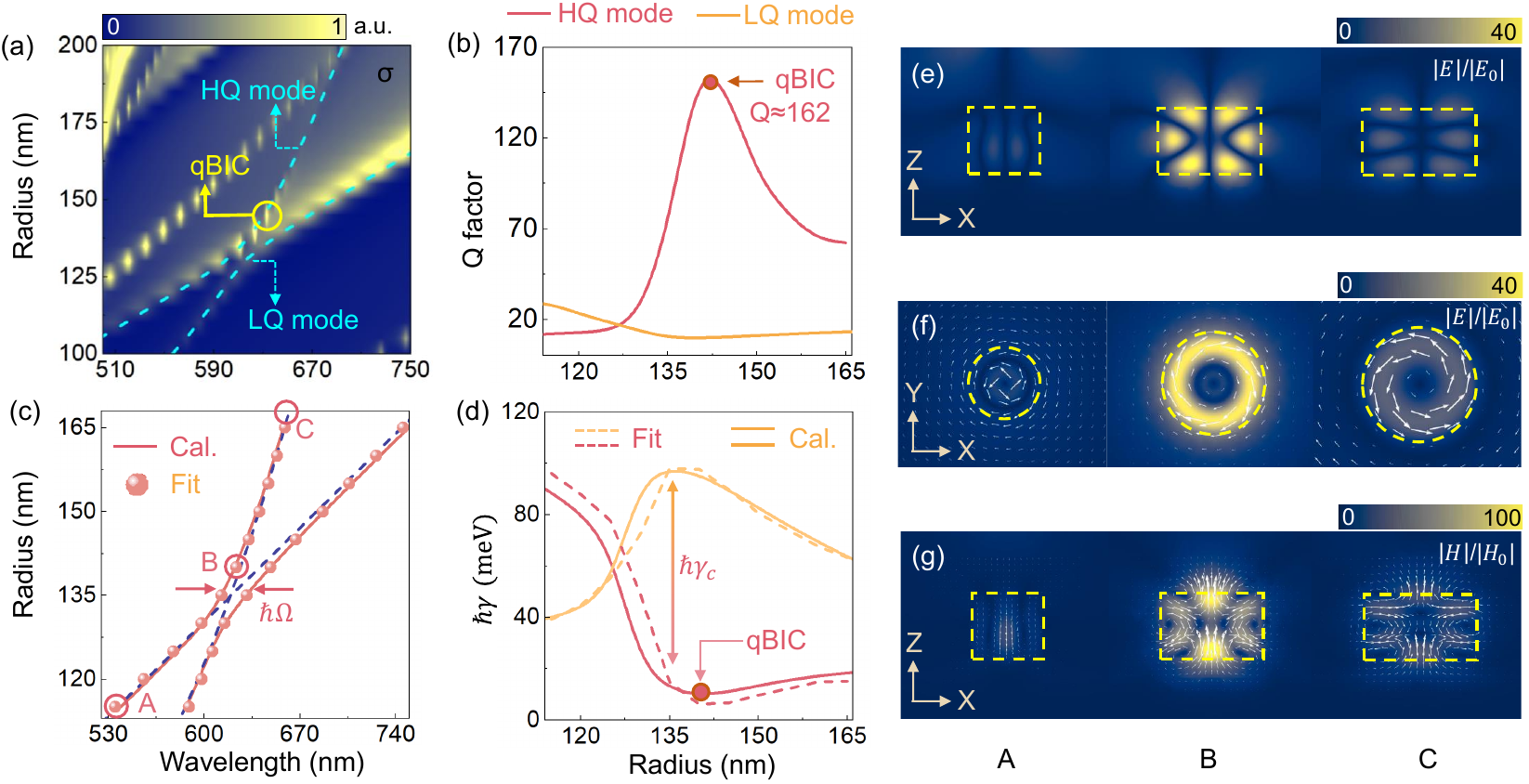}
  \caption{ Quasi-BIC in an isolated WS$_2$ nanodisk. (a) Optical modes in WS$_2$ nanodisk with background permittivity. Dashed cyan lines are the calculated high-Q and low-Q modes by the coupling oscillator model. (b) The Q factors versus the radius of nanodisk extracted from (a). (c) Fitted (dots) and calculated (solid lines) mode positions according to (a). The dashed lines marks the two vectorial Mie modes. (d) Fitted (dashed lines) and calculated (solid lines) damping energy of the two interfering optical modes. (e-g) The electric (e,f) and magnetic (g) field distribution within the nanoantennas. The white arrows indicate the corresponding direction of electric (f) and magnetic (g) field.}
  \label{fig2}
\end{figure*}

Figure~\ref{fig2}(a) presents the simulated scattering spectra as a function of nanodisk radius (100–200 nm), with a fixed nanodisk height of 200 nm. An anti-crossing behavior is clearly visible with the formation of two resonance branches (cyan lines), which arise from the interference of two vectorial Mie modes. Interestingly, the high-energy branch exhibits a narrow linewidth and a noticeable linewidth variation, whereas the low-energy branch exhibits invariant linewidth while varying the size of the nanodisks.  

To quantitatively analyze the spectral feature of both branches, we extracted their Q factors by fitting the scattering spectra (Fig.~\ref{fig2}a) to a modified Fano formula\cite{Melik2021NL}(see Supporting Information for fitting details). The extracted Q factors are given in Fig.~\ref{fig2}(b). It is evident that the low-energy branch maintains a consistently low Q-factor ($\sim$20) across all radii (red line), which is defined as low-Q (LQ) mode. In contrast, the Q-factor of the high-energy branch, marked as high-Q (HQ) mode, initially increases with increasing disk radius and reaches a maximum value of 160 at \( r = 140 \) nm (red dot). 

To quantitatively study the linear optical response of the coupled system, we then describe the system by an effective non-Hermitian Hamiltonian in the form of a 2$\times$2 matrix\cite{Xie2024NL}:
\begin{eqnarray}
H=\hbar\left[\left(\begin{array}{cc}	\omega_{1}-i\gamma_{1}	&	g		\\
                        		g		&	\omega_2-i\gamma_2\end{array}
\begin{array}{cc}     \\
                             \end{array}\right)
                             -i
                             \left(\begin{array}{cc}	0	&	\gamma_{c}		\\
                           		\gamma_{c}			&	0	\end{array}
\begin{array}{cc}     \\
                             \end{array}\right)\right]. \label{COM1}
\end{eqnarray}
Here, $\omega_i$ and  $\gamma_i$ are the angular frequencies and damping rates of the $i$th mode. $g$ represents the coherent coupling strength. Note that an important cross-damping term $\gamma_{c}$ is included to consider the influence of the coupling-induced incoherent interaction process. It is crucial to emphasize that incoherent coupling cannot be neglected in our case. We will show that the simulated optical characteristics of the hybrid system can be well reproduced provided that both coherent ($g$) and incoherent ($\gamma_{c}$) terms are taken into account. 

By solving Eq.~\ref{COM1} with optimized coupling energy $\hbar g$ = 30 meV and the cross-damping term $\hbar\gamma_{c}$= 36 meV, we excellently reproduce both the dispersion (dots in Fig.~\ref{fig2}c) and the spectral width (dashed lines in Fig.~\ref{fig2}d) of the two branches obtained by the above mentioned fitting procedure. More details for the calculation can be found in the Supporting Information (Section 2). In the present system, the coherent coupling energy $\hbar g$ described by a Hermitian matrix element $g$ exceeds $(\gamma_{1}+\gamma_{2})/2$, reaching the strong coupling regime\cite{Hertzog2019CSR}. In this case, the dipolar (Rabi) coupling can result in a coherent exchange of energy between them, oscillating
with a period of about $\pi/g$. In particular, we obtain distinctly different spectral linewidths for these two hybrid modes when bringing the two photonic Mie modes into resonance (arrow in Fig.~\ref{fig2}d). This indicates that in our system a vacuum-mediated incoherent dipole coupling leads to different radiative lifetimes of the LQ/HQ modes.

Essentially, these Mie modes can spontaneously emit photons into the surrounding vacuum field. These photons can then be reabsorbed, resulting in an incoherent exchange of energy between both optical oscillators, without any definite phase relationship. This process can be described by the non-Hermitian matrix element $\gamma_{c}$, the so-called cross-damping term\cite{Wang14ACSN,Xie2021PRB}. It is important that $\gamma_{c}$ and $g$ are not independent of each other but are related by a Kramers-Kronig relationship\cite{Akram00PRA}. On resonance (zero detuning), such an incoherent coupling process leads to short-lived (superradiant) and long-lived (subradiant) hybrid states with distinctly different decay rates. Specifically, the incoherent energy exchange between the two modes via the vacuum field fluctuations results in the subradiance effect of the high-Q mode, ultimately leading to the formation of the qBIC (red dot in Fig.~\ref{fig2}b,d)\cite{Rybin17PRL,Bogdanov19AP}. Such phenomena of sub- and superradiance are of great interest as they play a dominant role in the optical properties of many strongly coupled systems such as trapped ions\cite{DeVoe96PRL}, molecular aggregates\cite{Lim04PRL}, excitonic quantum dots\cite{Chen03PRL} and wells\cite{Hubner96PRL}, and plasmonic excitations in nanostructures\cite{Ropers05PRL,Martín10NL}.

We now investigate the self-hybridized interaction between the resonant modes and the excitons to explore the formation of polaritonic BICs. In this case, we now include the A exciton contribution (solid lines in Fig.~\ref{fig1}d). 

\begin{figure*}[t]
  \includegraphics[width=0.95\linewidth]{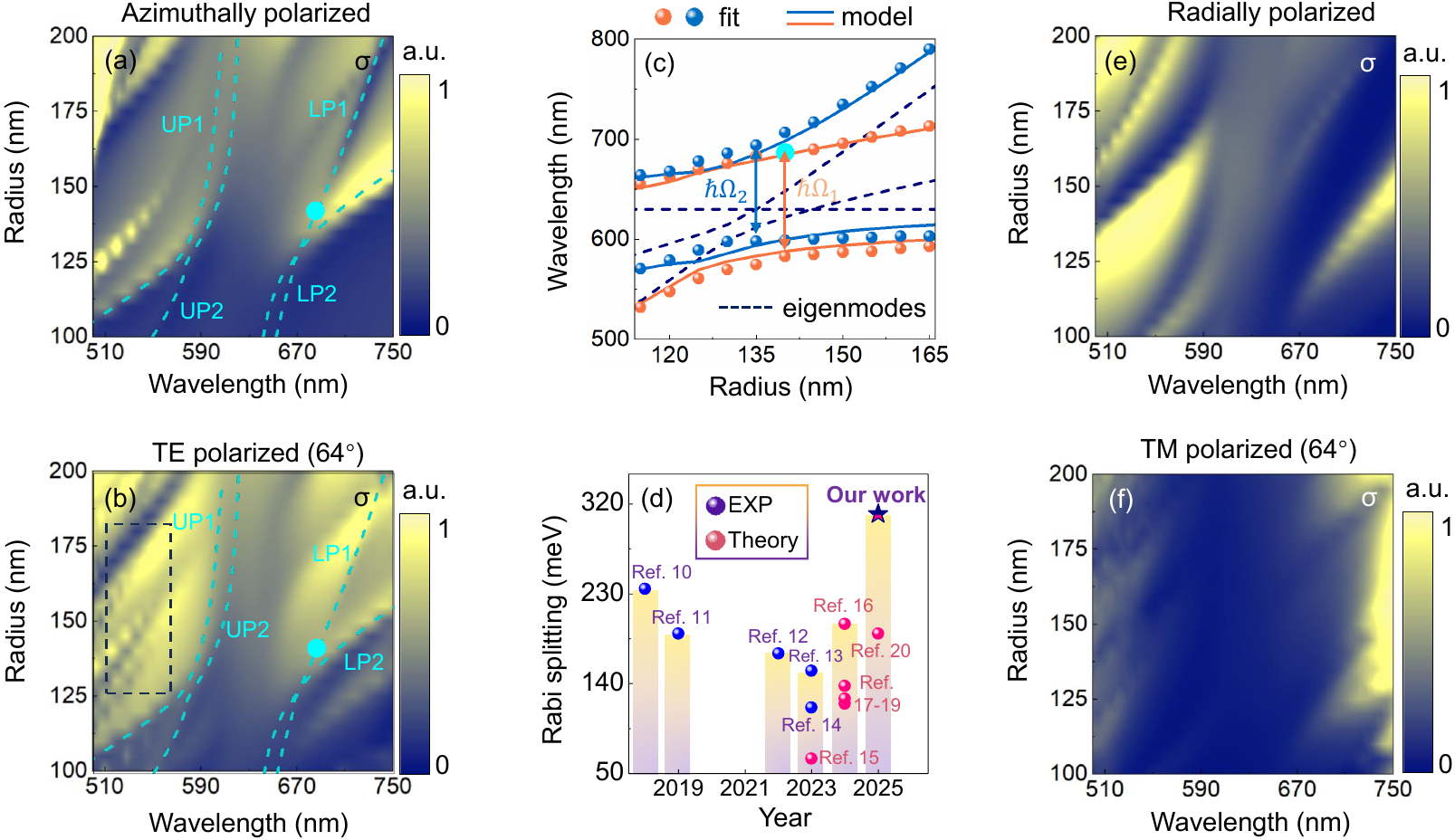}
  \caption{ Strong coupling between qBIC and excitons in WS$_2$ nanodisks.. (a,b,e,f) Simulated normalized scattering spectra of the WS$_2$ nanodisks for (a) azimuthal, (e) radial, (b) TE oblique and (f) TM oblique incidence. Dashed cyan lines in (a) and (b) present the calculated hybrid modes by coupling oscillator model.  (c) Fitted (dots) and calculated (solid lines) mode positions according to the scattering in (a). (d) Literature survey for Rabi splitting in all-dielectric self-hybridized TMDC nanostructures.}
  \label{fig3}
\end{figure*}

It is important to note that previous studies have demonstrated that FWBICs in a single dielectric nanoparticle can be excited under both azimuthally polarized light at normal incidence and TE-polarized light at oblique incidence\cite{Anastasiia23SA,Mylnikov20ACSN}. To compare the coupling behavior under these two excitation schemes, we simulated the scattering spectra as a function of the nanodisk radius for these two cases. The corresponding simulation results are shown in Figs.~\ref{fig3}(a) and \ref{fig3}(b), respectively. 

In both cases, four branches are observed in the scattering spectra, labeled as UP1, UP2, LP1, and LP2, respectively. All four branches are also clearly visible in the calculated absorption and extinction spectra (Section 3, Supporting Information). We emphasize that, regardless of whether vectorial light or linearly polarized light is used, the LQ mode and HQ mode couple separately with the A excitons, rather than forming a three-mode hybridization between two vectorial Mie modes and the exciton (which would exhibit three branches). Our conclusion is further validated by a coupled resonator model, which employs two 2×2 coupled Hamiltonians (Eq.~\ref{COM1}).  

Figure~\ref{fig3}(c) demonstrates that when the HQ-exciton and LQ-exciton coupling strengths were set to 150 meV and 140 meV respectively, the calculated dispersion relations (solid lines) align well with the simulated scattering spectrum obtained from numerical fitting (dots). In this case, the corresponding Rabi splittings reach $\hbar\Omega_{1}=310$ meV and $\hbar\Omega_{2}=280$ meV, respectively, with the formation of a polaritonic qBIC on HQ-exciton hybrid branch (the cyan dot). We overlaid the calculated dispersion relations (dashed cyan lines) onto the simulated spectra in Figs.~\ref{fig3}(a) and \ref{fig3}(b), confirming that both excitation schemes exhibit the coupling behavior predicted by theory. Furthermore, in this system, the damping rates of the HQ mode and LQ mode at zero detuning are 6 meV and 98 meV, respectively, while the damping rate of the A exciton of WS$_2$ is 51 meV, indicating that both coupled subsystems reach the strong coupling regime. We attribute the giant energy splitting observed in the coupling of the HQ mode and the exciton to the strong electromagnetic field confinement within the nanoresonator. Figures~\ref{fig2}(e-g) show the near-field distributions for the nanodisks with different radii, as marked by A, B, C in Fig.~\ref{fig2}(c). The electric field distribution exhibits typical qBIC characteristics, as demonstrated in the middle panels of Figs.~\ref{fig2}(e-g). Significantly, this qBIC mode arises from the interference between two in-plane electric field-dominated Mie resonances (denoted as Point A and Point C), generating a remarkable ring-like in-plane electric field enhancement. The maximum electric field enhancement factor is above 40, surpassing the localized field intensities reported in most dielectric nanoantennas to date. Notably, this magnetic octupole-dominated qBIC mode simultaneously achieves an exceptional magnetic field enhancement factor exceeding 100, as evidenced in the middle panel of Fig. 3(g). The strong in-plane electric field localization and its distinctive spatial distribution substantially improve the spatial mode matching with in-plane excitons, thereby dramatically enhancing the light-matter coupling efficiency. Due to the exceptional field confinement capability of the qBIC, the HQ mode hybridized with the exciton reaches the largest Rabi splitting observed in all-TMDC self-hybridized systems (Fig.~\ref{fig3}d).

Notably, the difference between the azimuthal and TE excitation schemes lies in the symmetry of the system. Under TE-polarized excitation, the system exhibits lower symmetry, leading to the presence of additional resonance modes (dashed-line area in Fig. 3(b)). The emergence of these extra modes interferes with the identification of characteristic modes in coupled systems, as well as the observation and analysis of coupling behavior.  

Furthermore, under radially polarized and TM-polarized excitations, which are orthogonal to the two previously studied polarization schemes, the optical modes supported by the nanodisk do not form qBIC near the WS$_2$ exciton resonance (Section 4, Supplementary Information). This enables selective excitation of polaritonic BICs. To verify this, we simulated the scattering spectra under radially polarized and TM linearly polarized excitation, as shown in Figs.~\ref{fig3}(e) and \ref{fig3}(f). As expected, similar to the azimuthal and TE-polarized cases, the TM-polarization case exhibits more complex resonance modes without the formation of polaritonic qBIC, but the optical responses under radial and TM polarizations remain similar. This behavior is consistent with our theoretical predictions.


\begin{figure*}[t]
  \includegraphics[width=0.95\linewidth]{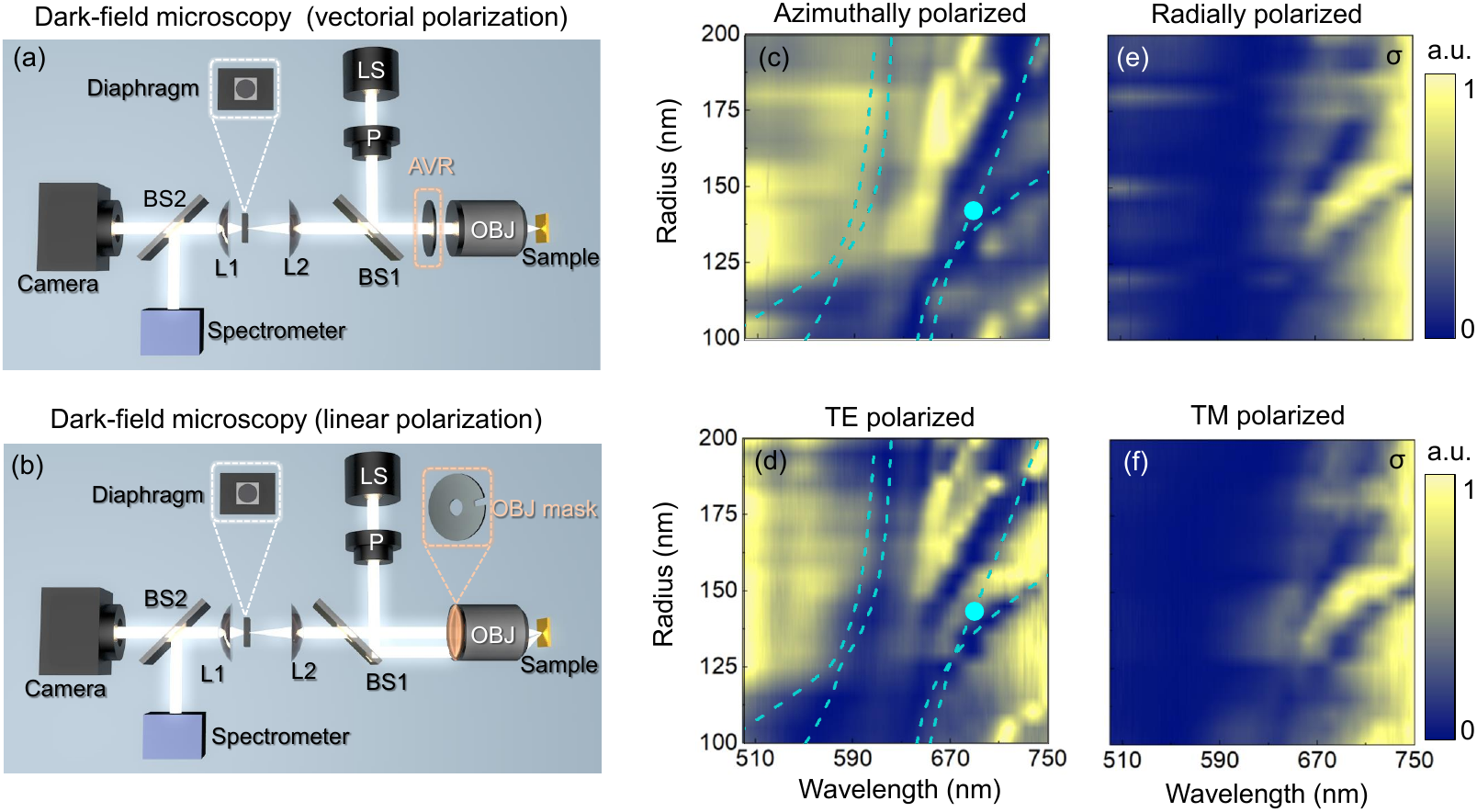}
  \caption{ Experimental measurement of scattered field. (a,b) The experimental configuration of scattering measurement for (a) vectorial and (b) linear polarization, respectively. (c-f) Measured normalized scattering spectra of WS$_2$ nanodisk for (c) azimuthal, (e) radial, (d) TE oblique and (f) TM oblique incidence. Dashed cyan lines in (a) and (b) indicate the results of theoretical calculations. }
  \label{fig4}
\end{figure*}


To validate the theoretical analysis, dark-field spectroscopy was utilized to characterize the optical response of the fabricated WS$_2$ nanodisk samples. We used two types of microscopic setup to either excite the samples with tightly focused vectorial polarized light or linearly polarized light at oblique incidence, as shown in Fig.~\ref{fig4}(a,b). For vectorial excitation, an achromatic vortex retarder (AVR) was placed in front of the objective to generate azimuthally/radially polarized light, and the scattered light from the sample was collected collinearly with the incident beam (Fig.~\ref{fig4}(a)). For the linear polarization case, the dark-field setup was constructed for the illumination of a segment of the input outer ring of the dark-field microscope objective (see Fig. S7). As shown in Fig.~\ref{fig4}(b), an objective mask with a side slit was placed behind the objective. In this case, the incident beam was guided to pass through the side slit so that the sample was illuminated at oblique incidence. The light was then scattered back by the nanodisk and collected from the center of the objective (see Methods for details).

The measured scattering spectra of WS$_2$ nanodisks under azimuthally polarized and TE-polarized light excitation are given in Figs.~\ref{fig4}(c) and \ref{fig4}(d), respectively. Remarkably, the principal spectral features of the two sets of spectra demonstrate great consistency, particularly in terms of peak positions and relative intensities, which is in excellent agreement with our theoretical analysis (Fig.~\ref{fig3}a,b). The theoretical dispersion relations derived from the coupled-resonator model (cyan curves) were overlaid onto the measured scattering spectra, revealing a quantitative agreement between measured spectral features and calculated eigenmode solutions. 

Similarly, we performed measurements using radially polarized and TM-polarized light excitation, with the resulting scattering spectra shown in Figs.~\ref{fig4}(e) and \ref{fig4}(f), respectively. In these two cases, no significant coupling behavior was observed. According to the simulation results, in the short-wavelength region, the resonance modes either exhibit excessive losses (Fig.~\ref{fig4}e) or possess highly complex optical characteristics with weak resonance strength (Fig.~\ref{fig4}f). These factors prevent the observation of short-wavelength resonance modes. Consequently, we can easily distinguish whether a polaritonic qBIC is excited on the basis of the experimental data.


Therefore, both theoretical and experimental results confirm the excitation of polaritonic BICs under azimuthally polarized and TE-polarized light. The hybridization between excitons and qBICs can be observed more clearly under azimuthally polarization. Additionally, our findings demonstrate the feasibility of observing self-hybridization behavior under TE-polarized excitation, providing a more convenient approach for the experimental investigation of polaritonic BICs.

\section{Conclusions}
 
In summary, we have presented the first theoretical analysis and experimental realization of polaritonic quasi-bound states in the continuum within isolated WS$_2$ nanodisks through intrinsic coupling between Mie-type photonic resonances and in-plane exciton transitions. By implementing azimuthally polarized and TE-polarized excitation configurations, we have characterized the hybrid states using polarization-controlled scattering spectroscopy, thereby unambiguously confirming the existence of polaritonic qBIC. The system achieves an extraordinary Rabi splitting energy of 310 meV, representing a great enhancement compared to previous records in all-TMDC systems. This record-breaking coupling strength originates from the unique synergy between qBIC-induced electromagnetic field confinement and exciton oscillator strength – specifically, the in-plane electric field distribution spatially overlaps with excitonic dipoles while the non-radiative nature of qBIC modes suppresses photon leakage. Polarization-controlled measurements reveal that switching between orthogonal polarizations enables selective excitation of polaritonic qBIC. Such polarization-selective control provides unprecedented spatial and spectral manipulation of polaritonic dispersion without structural modifications. Our findings open avenues for developing room-temperature polariton lasers, nonlinear optical switches, and quantum emitters with enhanced light-matter interactions.

\section{Methods}

\subsection{Sample fabrication}
Single-crystalline WS$_2$ flakes were mechanically exfoliated from bulk crystals (HQ Graphene) and transferred onto substrates comprising a 100 nm-thick gold layer and a 55 nm thick SiO$_2$ layer. Then a negative resist layer (ma-N 2403, micro resist) was spin coated and soft baked prior to patterning. Electron-beam lithography was performed at 20 kV with a dose of 120 µC/cm$^2$ to define the desired structures. The samples were subsequently subjected to reactive-ion etching to remove the unprotected WS$_2$. Finally, the remaining resist was dissolved using a commercial remover, followed by a thorough rinse in water.

\subsection{Numerical simulation}
The numerical simulations were conducted in the frequency domain using COMSOL Multiphysics software. All computations were performed for a single nanoresonator of a fixed size (200 nm in height) positioned on a multilayered substrate and enclosed by a perfectly matched layer to simulate an infinite surrounding medium. The material properties, including losses, were obtained from tabulated data for Au in the visible to near-infrared range\cite{Rakic98AO} and from experimental ellipsometry measurements for bulk WS$_2$. Discrepancies between the simulated and measured resonant peak positions are likely due to fabrication imperfections, such as the inclination of the disk walls, which were not incorporated into the simulation model.

\subsection{Optical characterization}
To perform optical measurements, we designed two experimental setups: vectorial excitation and linear excitation. For vectorial excitation, a liquid crystal polymer-based achromatic vortex retarder (Lbtek) was used to generate a vectorial pump beam from a linearly polarized Gaussian white light source (Olympus TH4-200). The pump beam was visualized using a camera (Hikrobot MV-CS050-10UC) in combination with an achromatic lens. The sample was mounted on a three-dimensional translation stage, and the pump beam was focused onto it from the front side using a Olympus objective lens (MPlanFL N,100×, 0.90 N.A., BD) installed on an Olympus dark field microscope system (BX53M) to achieve a tightly focused beam waist. In the working spectral range, the gold layer acted as a reflective mirror. The reflected signal was collected in the backward direction by the same objective lens (0.90 N.A.), passed through a non-polarizing beamsplitter, and was subsequently detected using a CCD (Andor iVac 316) connected to the spectrometer (Zolix Omni-$\lambda$5028i). To obtain the backward scattering spectra, the power spectrum of the reflection from a single nanoparticle was normalized by the power spectrum of the reflection from the substrate.

For linear excitation, a linearly polarized Gaussian beam was illuminating a segment of the outer ring of the dark-field microscope objective (MPlanFL N, 100$\times$, 0.90NA BD). Supplementary Figure S7 shows a simplified schematics of the illumination. An incident linear polarization with the electric field oriented towards/away from the lens centre thus resulted in the oblique TM-polarised excitation. The electric field oriented perpendicular to the direction of the lens centre resulted TE-polarized excitation. The scattered signal was collected with the central part of the same objective lens. Signal was detected with a spectrometer Ocean Optics QEPro and a camera Trius-SX694 Starlight Xpress camera paired with an infinity-corrected f $=$ 150 mm visible achromatic doublet lens.

\begin{acknowledgements}

This work was supported by the Natural Science Foundation of Sichuan Province in China (Grant No. 2024NSFSC0460), Australian Research Council (grants DP210101292 and DE210100679), International Technology Center Indo-Pacific (ITC IPAC) via Army Research Office (Contract FA520923C0023), and the European Union (ERC, METANEXT, 101078018). Views and opinions expressed are however those of the author(s) only and do not necessarily reflect those of the European Union or the European Research Council Executive Agency. Neither the European Union nor the granting authority can be held responsible for them. This project was also funded by the Deutsche Forschungsgemeinschaft (DFG, German Research Foundation) under grant numbers EXC 2089/1–390776260 (Germany’s Excellence Strategy) and TI 1063/1 (Emmy Noether Program), the Bavarian program Solar Energies Go Hybrid (SolTech) and the Center for NanoScience (CeNS).

\end{acknowledgements}

\newpage

 \bibliography{References.bib}


\end{document}


\maketitle

\clearpage

\noindent\textbf{1. Sample fabrication}

\begin{figure*}[!htb]

	\centering
	\includegraphics[width=12 cm]{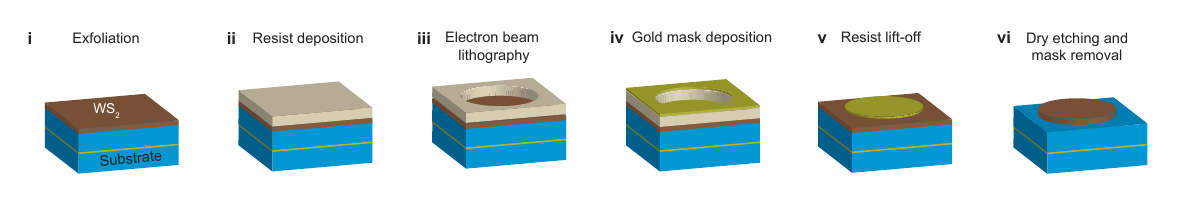}
	\caption{ Process for sample preparation.}
\end{figure*}

\begin{figure*}[!htb]
	\centering
	\includegraphics[width=12 cm]{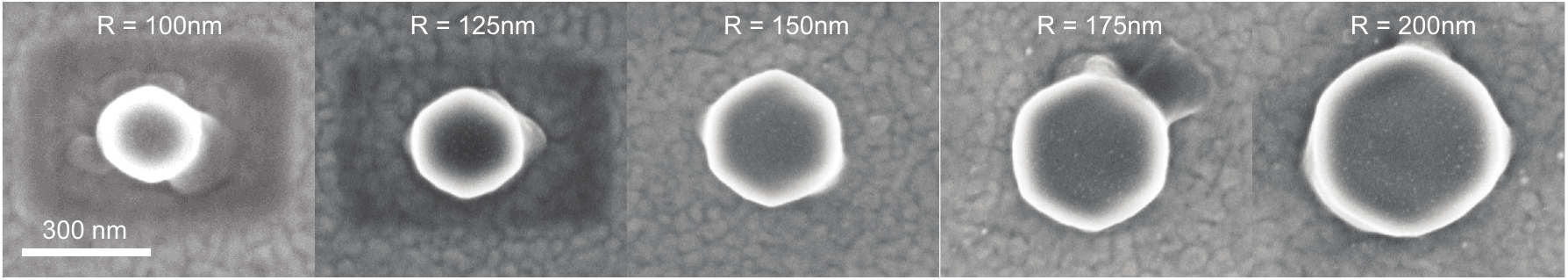}
	\caption{ Scanning electron microscopy (SEM) top-view images showing nanodisk arrays with radii in ascending order: 100 nm, 125 nm, 150 nm, 175 nm, and 200 nm.}
\end{figure*}

\clearpage

\noindent\textbf{2. Vectorial Mie modes}

\begin{figure*}[!htb]
	\centering
	\includegraphics[width=8 cm]{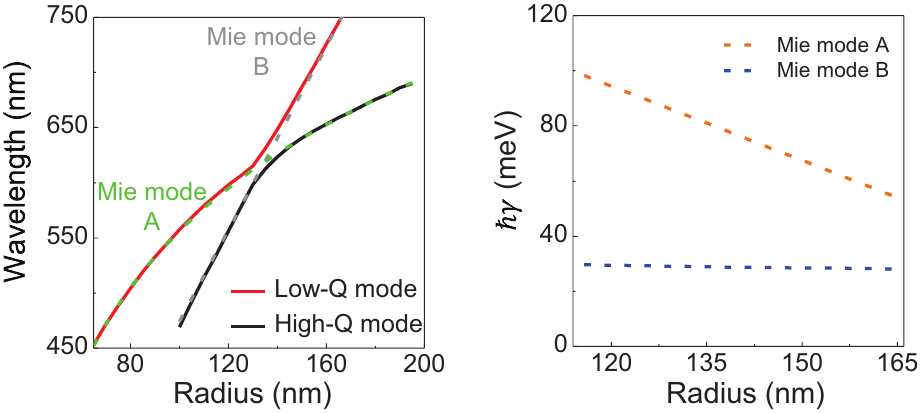}
	\caption{ (a) The resonant wavelengths of eigen Mie modes, high-Q mode and low-Q mode. (b) The damping rates of the two Mie modes.}
\end{figure*}

For Fano resonance in our work, the scattering cross-section $\sigma$ can be described by the modified Fano formula \cite{Andrey19AP}:
\begin{eqnarray}\label{Sfit}
\sigma(\omega)=\sigma_{bg}+\sum_{j=1}^N\frac{\sigma_j}{1+q_j^2}\frac{(q_j+x_j)^2}{1+x_j^2},
\end{eqnarray}
here $x_j = 2(\omega - \omega_j)/\gamma_j$ donates the normalized frequency offset of $j$th mode, with $\omega$, $\omega_j$ and $\gamma_j$ representing the frequency of photon, the resonance frequency and the inverse of lifetime of $j$th mode. The Fano asymmetry parameter $q$, as the key parameter that determines the qBIC with a maximum Q factor \cite{Melik21NL}, can be calculated by $q_j = -cot(\delta_j)$, where $\delta_j$ is the phase shift between the $j$th resonant mode and the background contribution \cite{Alpeggiani17PRX}. $\sigma_j$ and $\sigma_{bg}$ are the smooth peak envelope of $j$th mode and the background contribution, which both originate from the nonresonant contribution of other modes.

In our 2×2 Hamiltonian, to obtain the dispersion and damping rates of the eigen Mie modes,we approximate the behavior of the eigenmodes by fitting the spectra of the Mie modes at detunings far from zero. We assume that the resonance wavelengths of Mie mode A (green dashed line) and Mie mode B (gray dashed line) are as shown in Figure (a). Additionally, we assume that the linewidth of the Mie modes varies linearly with the nanodisk radius, resulting in the damping rates shown in Figure (b).

\clearpage

\noindent\textbf{3. Scattering, absorption and extinction spectra of nanodisks.}

\begin{figure*}[!htb]
	\centering
	\includegraphics[width=12 cm]{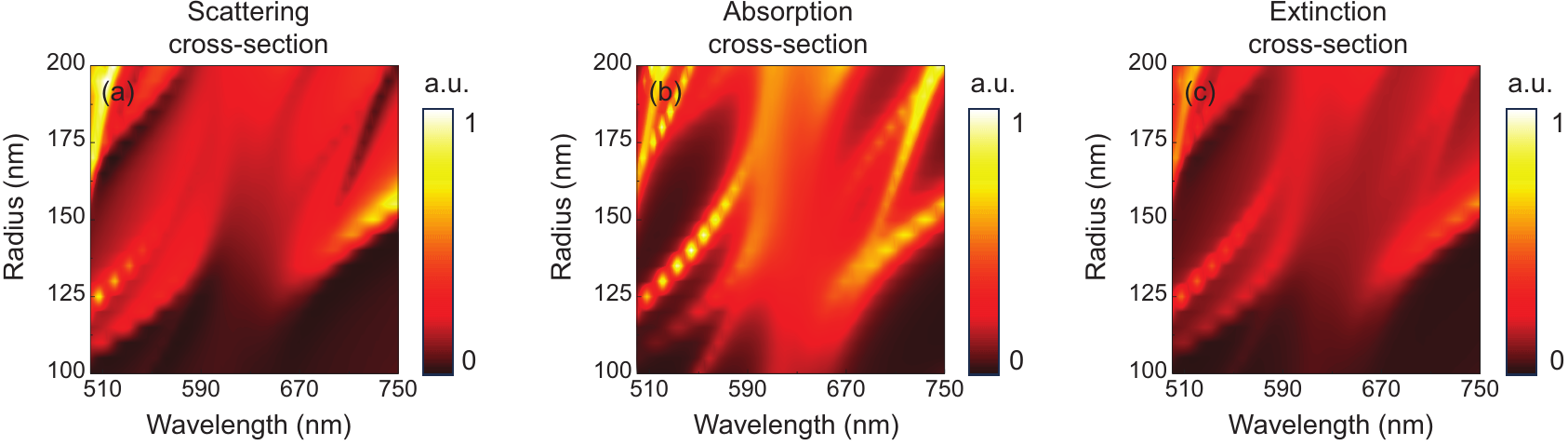}
	\caption{ Simulated (a) scattering, (b) absorption and (c) extinction spectra of the nanodisk under azimuthally polarized excitation.}
\end{figure*}

\clearpage

\noindent\textbf{4. Permittivity of bulk WS$_2$ and eigen optical response of nanodisks}

\begin{figure*}[!htb]
	\centering
	\includegraphics[width=12 cm]{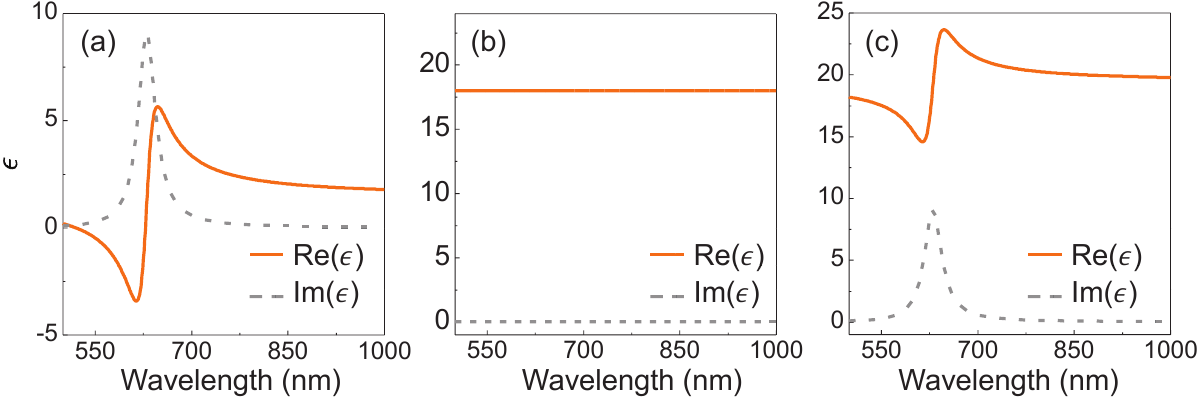}
	\caption{ Artificial permittivities used to simulate scattering from nanodisks made of a 
hypothetical (a) exciton-only nanodisk, (b) lossless high-index nanodisk (blue line), and 
(c) sum of these permittivities (red lines). Solid lines show real parts, and dashed lines are imaginary parts.  }
\label{figS5}
\end{figure*}

\begin{figure*}[!htb]
	\centering
	\includegraphics[width=12 cm]{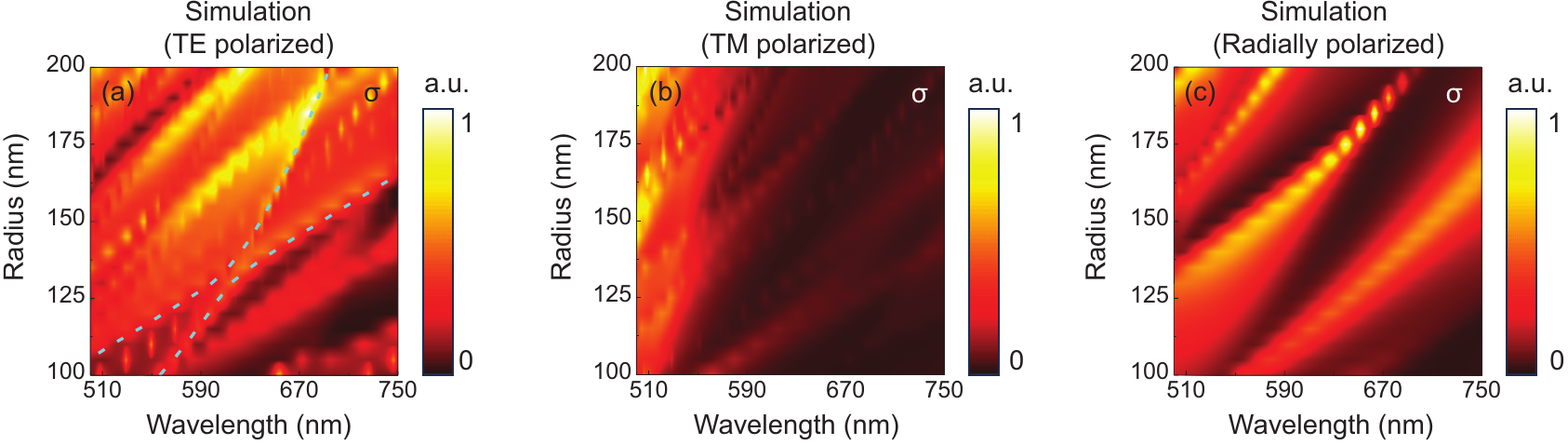}
	\caption{ Simulated scattering spectra of nanodisk with the permittivity from fig. \ref{figS5}(b) for (a) TE oblique, (b) TM oblique and (c) radial polarization.}
	\label{figS6}
\end{figure*}

The artificial permittivity of bulk WS$_2$ was described by a sum of Lorentzian oscillators \cite{Wang16NL, Li14PRB, Xie21PRB}:
\begin{eqnarray}
\varepsilon(E)=\varepsilon_B+\sum_{j=1}^N\frac{f_j}{E_j^2-E^2-iE\Gamma_j},\label{LorentzOscillation}
\end{eqnarray}
Here, for the permittivity from fig.~\ref{figS5}(c), $E$ and $\varepsilon_B = 19$ donate the photon energy and the background permittivity. $f_1 = 1.836$, $E_1 = 1.968 eV$ and $\Gamma_1=2\gamma_1 = 0.1026 eV$ are the oscillator strength, resonance energy and the full linewidth of $j$th order oscillator where $\gamma_1$ is the damping rate. For simplicity, as show in fig.~\ref{figS5}(b), we assume the  bulk WS$_2$ as a dispersionless material with permittivity $\varepsilon = 18 + 0.01i$ to study the eigen optical responses of nanodisks in our simulation (fig.~\ref{figS6}).

\noindent\textbf{5. Schematic of illumination configuration for the linear TE/TM measurements}

\begin{figure*}[!htb]

	\centering
	\includegraphics[width=12 cm]{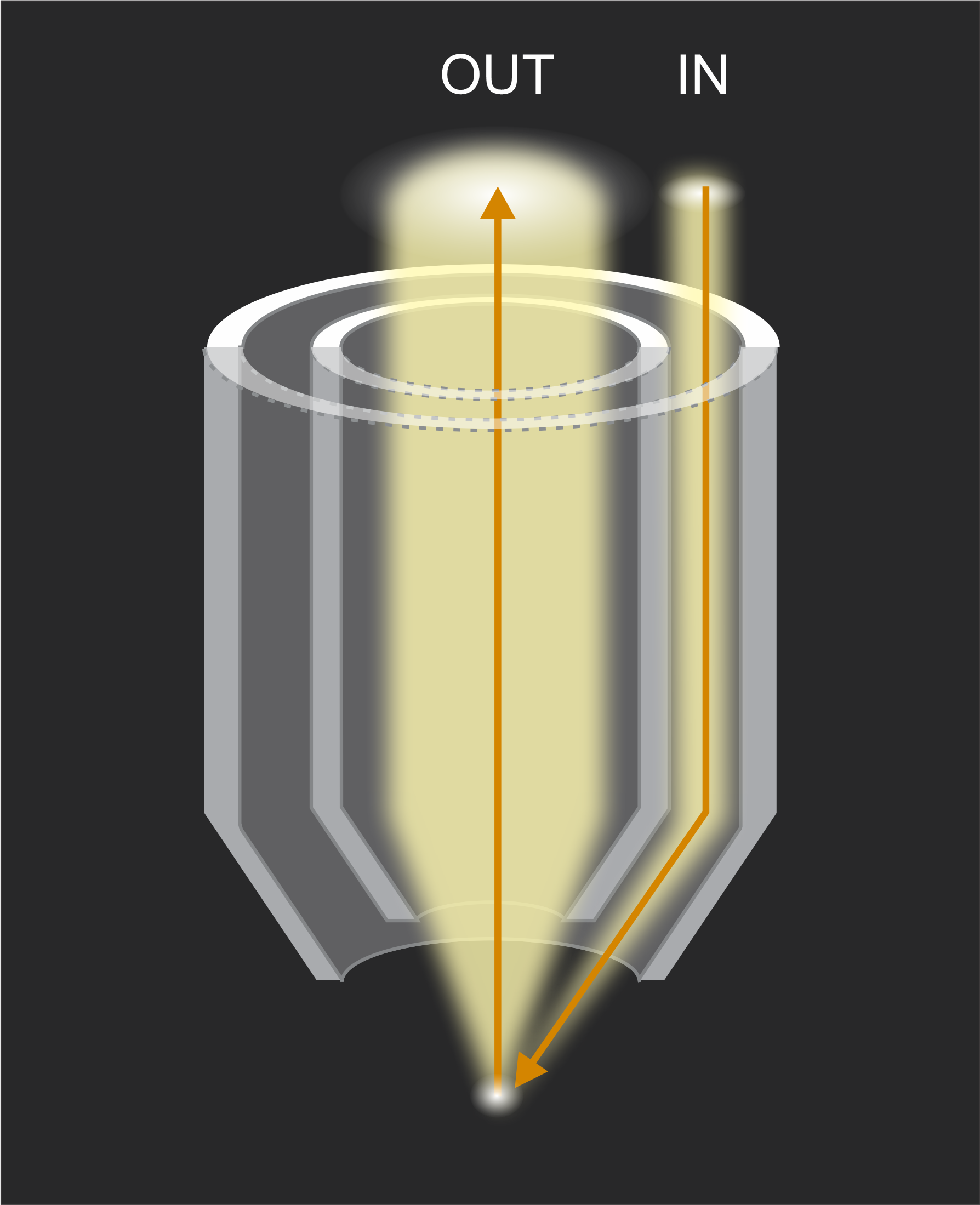}
	\caption{Simplified sketch of the illumination setting for the TE/TM linear polarization measurements. Linearly polarized white light was illuminating a segment of an outer ring of a dark field microscope objective resulting in the oblique TE/TM incidence.}
\end{figure*}

\clearpage

\bibliography{main}